\documentclass{webofc}
\usepackage[varg]{txfonts}   
\usepackage{hyperref}
\begin{document}
\title{Dynamic modeling for heavy-ion collisions}

\author{\firstname{Chun} \lastname{Shen}\inst{1,2}\fnsep\thanks{\email{chunshen@wayne.edu}}}

\institute{
Department of Physics and Astronomy, Wayne State University, Detroit, Michigan, 48201, USA 
\and
RIKEN BNL Research Center, Brookhaven National Laboratory, Upton, NY 11973, USA}

\abstract{%
Recent theory progress in (3+1)D dynamical descriptions of relativistic nuclear collisions at finite baryon density are reviewed. Heavy-ion collisions at different collision energies produce strongly coupled nuclear matter to probe the phase structure of Quantum Chromodynamics (QCD). Dynamical frameworks serve as a quantitative tool to study properties of hot QCD matter and map collisions to the QCD phase diagram. Outstanding challenges are highlighted when confronting theoretical models with the current and forthcoming experimental measurements from the RHIC beam energy scan program.
}
\maketitle
\section{Introduction}
\label{intro}

Quantifying the phase structure of hot and dense QCD matter is one of the primary goals in relativistic nuclear physics. Lattice QCD calculations \cite{Bhattacharya:2014ara, Bazavov:2011nk} provide conclusive information that hot nuclear matter at zero net baryon density transits from the hadron gas to the Quark-Gluon Plasma (QGP) phase via a smooth crossover at the pseudo-critical temperature $T_{pc} = 156.5 \pm 1.5$\,MeV \cite{HotQCD:2018pds}. Moreover, the same type of phase structure remains for small-to-moderate values of the net baryon chemical potential $\mu_B$ \cite{HotQCD:2018pds}.
It was conjectured that the quark-hadron transition turns from a crossover to a first-order phase transition at some finite baryon chemical potential, suggesting the existence of a critical point \cite{Asakawa:1989bq, Bzdak:2019pkr}. However, the first principles lattice QCD calculations can only provide limited guidance on the existence and location of the QCD critical point in the $T - \mu_B$ phase diagram because of the fermion sign problem at finite net baryon density \cite{deForcrand:2009zkb}.
Meanwhile, exploration of the dense quark matter properties at large net baryon density is of particular importance since the gravitational waves detection, emerging from black hole/neutron star mergers, now give stringent constraints on the properties of the compact stars, including the nuclear matter equation of state (EoS) \cite{LIGOScientific:2017vwq, LIGOScientific:2018cki, Baym:2017whm, Dexheimer:2020zzs}.

To establish quantitative connections between the QCD phase structure and measurements from relativistic heavy-ion collisions over an extensive collision energy range, we need to model the entire dynamical evolution of the heavy-ion collisions. 
Relativistic hydrodynamics, incorporated with nuclear matter EoS, viscosity, and initial state fluctuations, has been a precision tool to understand the macroscopic dynamics of the strongly coupled QGP. The fluid dynamic description is connected with hadronic transport approaches to microscopically describe the system's evolution in the dilute phase. Such a hydrodynamics + hadronic transport hybrid theoretical framework has successfully described and even predicted various flow correlation measurements with remarkable precision \cite{Shen:2014vra, Bernhard:2016tnd, Schenke:2020mbo, JETSCAPE:2020mzn}. Phenomenological studies with precise flow measurements of the hadronic final state from the Beam Energy Scan (BES) program can elucidate the collective aspects of the baryon-rich QGP and constrain the QGP EoS and transport properties, such as its viscosity and charge diffusion coefficients \cite{Denicol:2018wdp, McLaughlin:2021dph}.
Strange quarks produced in relativistic nuclear collisions are an interesting probe for studying the evolution of the collisions. Since the collision system is strangeness neutral, the $s\bar{s}$ pair production mechanism is sensitive to the properties of strongly-interacting matter, especially at the early time \cite{Pratt:2018ebf}. The partonic and hadronic production channels are very different and may signal the onset of deconfinement and the quark-gluon plasma \cite{Blume:2017icv}. Furthermore, unstable resonance states of strange baryons can provide detailed information about hadronic reactions in the late-stage of heavy-ion collisions \cite{Oliinychenko:2021enj}. This proceeding will highlight the recent developments of the hybrid dynamical frameworks, emphasizing the strangeness-related observables.

\section{Flowing through the QCD crossover region}

Relativistic heavy-ion collisions with collision energies $\sqrt{s_\mathrm{NN}} \gtrsim 20$\,GeV can probe the nuclear matter properties with a net baryon chemical potential up to $\sim 300$\,MeV, where the QGP and hadronic phases are connected with a smooth crossover \cite{Andronic:2017pug, Adamczyk:2017iwn}. At $\sqrt{s_\mathrm{NN}} \sim \mathcal{O}(10)$\,GeV, heavy-ion collisions violate longitudinal boost invariance, and the overlapping time for the two nuclei to pass through each other is significant compared to the total collision lifetime \cite{Karpenko:2015xea, Shen:2017bsr, Shen:2020jwv}. Therefore, it is crucial to develop initial conditions with non-trivial longitudinal dynamics. The complex 3D collision dynamics can be approximated by parametric energy depositions based on the collision geometry \cite{Hirano:2005xf, Bozek:2017qir, Shen:2020jwv, Sakai:2020pjw}. Non-trivial dynamics could be included by modeling the energy loss in individual nucleon-nucleon collisions based on either string deceleration \cite{Shen:2017bsr, Bialas:2016epd} or full transport simulations \cite{Pang:2012he, Karpenko:2015xea, Du:2018mpf}. Further theory developments exist to understand the early-stage baryon stopping from the Color Glass Condensate-based approaches in the fragmentation region \cite{Li:2018ini, McLerran:2018avb} and a holographic approach at intermediate couplings \cite{Attems:2018gou}. In the meantime, dynamical initialization schemes have been developed to interweave initial-state and hydrodynamics on a local basis to model the extended interaction region between the two colliding nuclei at low collision energies \cite{Shen:2017ruz, Shen:2017bsr,Du:2018mpf, Akamatsu:2018olk}. These schemes have also been applied to study small systems and jet-medium interactions \cite{Okai:2017ofp, Kanakubo:2019ogh}.

\begin{figure}[t!]
\centering
\includegraphics[width=0.83\linewidth]{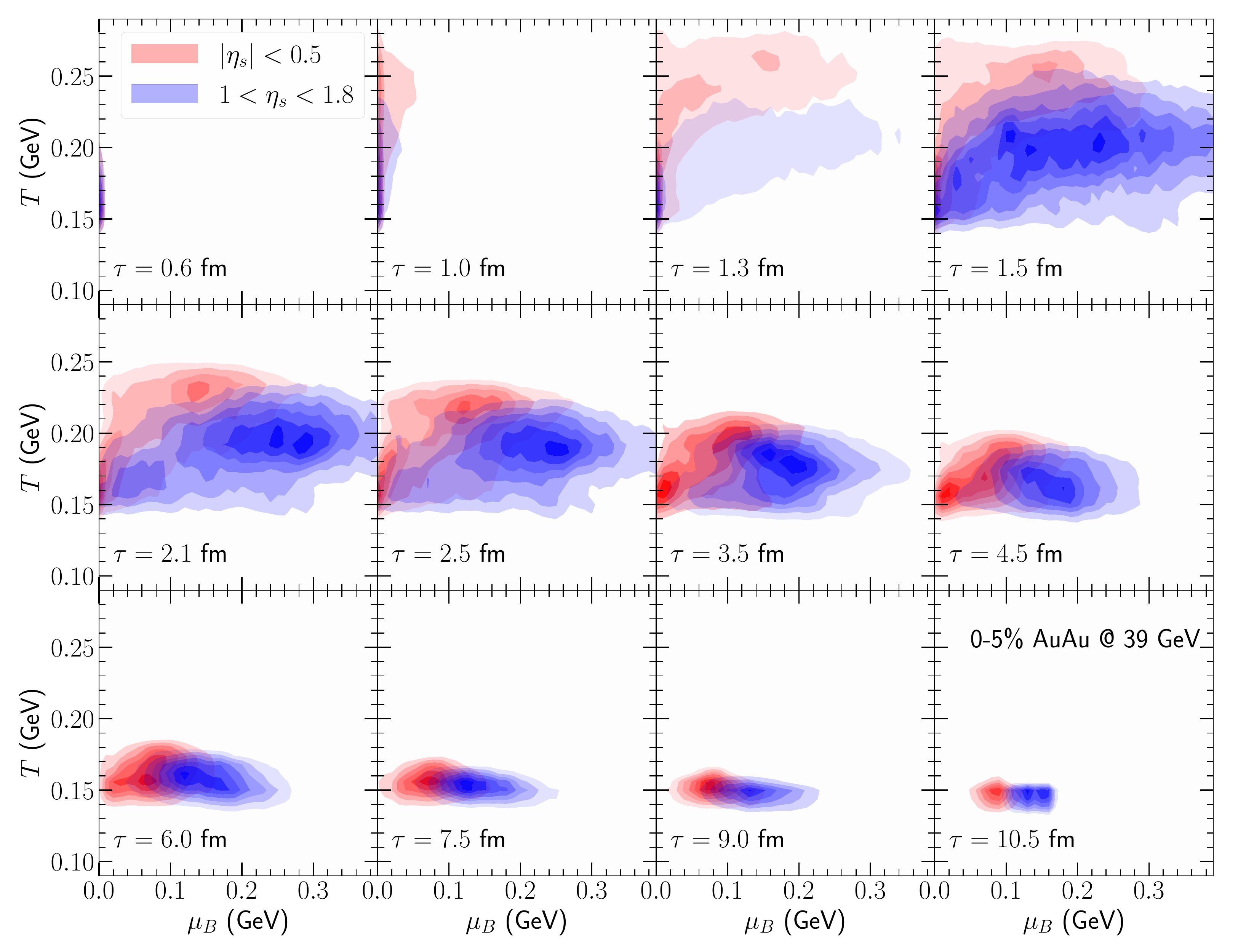}
\caption{The QCD phase diagram is dynamically probed by a central Au+Au collision at $\sqrt{s_\mathrm{NN}} = 39$\,GeV. The red and blue contours indicate the time snapshots of fluid cell $T-\mu_B$ distributions in the mid-rapidity $\vert \eta_s \vert < 0.5$ and a forward space-time rapidity region $1 < \eta_s < 1.8$, respectively.}
\label{fig:PhaseTrajectory}
\end{figure}
Figure~\ref{fig:PhaseTrajectory} shows a series of snapshots of one central Au+Au collision at $\sqrt{s_\mathrm{NN}} = 39$\,GeV flowing through the QCD phase diagram. The numerical simulation is based on 3D MC-Glauber with string deceleration model \cite{Shen:2017bsr}. The first row presents the early dynamical initialization stage $\tau \le 1.5$ fm/$c$, during which the two colliding nuclei pass through each other. The collisions among nucleons continuously deposit energy-momentum currents to form the strongly coupled QGP. Because most of the baryon charge sources are at the end of strings, their deposition times are later than those of the energy-momentum currents. Hence, the hydrodynamic medium first heats up at $\mu_B \sim 0$. Around $\tau \sim 1.2$ fm/$c$, the baryon charges start to be doped into the system and drive the system to flow to a large $\mu_B$ region in the phase diagram. For central collisions at $\sqrt{s_\mathrm{NN}} = 39$\,GeV, the duration of energy-momentum and net baryon charge depositions last for about 1.5-2 fm/$c$. During this phase, the fireball reaches $T_\mathrm{max} \sim 250$\,MeV and $\mu_{B, \mathrm{max}} \sim 200$\,MeV at the mid-rapidity. For fireball in the forward space-time rapidity region $1 < \eta_s < 1.8$, the peak temperature of the medium drops to about 200 MeV while the medium reaches out to a larger $\mu_B$ region, $\mu_B \sim 300$\,MeV.
For $\tau \ge 2$ fm/$c$, the fireball expands hydrodynamically in 3D and lives up to $10.5$\,fm/$c$ before all the fluid cells convert to hadrons in the dilute hadronic phase.

Because most hybrid dynamical frameworks assume grand-canonical ensemble (GCE) at the particlization, the hadron chemistry is mainly controlled by the distributions of temperature and chemical potentials on the constant energy density particlization hyper-surface. Furthermore, the late-stage hadronic transport model can further modify the relative particle abundances via inelastic scatterings and baryon-anti-baryon annihilation. Therefore, the yield ratios of identified particles provide strong constraints on the particlization condition in hybrid simulations \cite{Monnai:2019hkn, Alba:2020jir}. Assuming GCE, one can map the net electric charge $(Q)$, strangeness $(S)$, and baryon $(B)$ chemical potentials to identified particle ratios as follows, $\pi^-/\pi^+ \propto \exp(-2\mu_Q)$, $K^-/K^+ \propto \exp[-2 (\mu_Q + \mu_S)]$, and $\bar{p}/p \propto \exp[-2(\mu_Q + \mu_B)]$. The ratios $K^-/\pi^- \propto \exp(-\mu_S)$ and $K^+/\pi^+ \propto \exp(+\mu_S)$ provide direct information about the strangeness chemical potential. The yield ratios of multi-strange baryons to their antiparticles $\bar{\Lambda}/\Lambda \propto \exp[-2(\mu_B - \mu_S)]$, $\bar{\Xi}^+/\Xi^- \propto \exp[-2 (\mu_B - 2 \mu_S - \mu_Q)]$, and $\bar{\Omega}/\Omega \propto \exp[-2 (\mu_B - 3\mu_S)]$ contain different weights on $\mu_S$. 
\begin{figure}[ht!]
\centering
\begin{tabular}{cc}
   \includegraphics[width=0.4\linewidth]{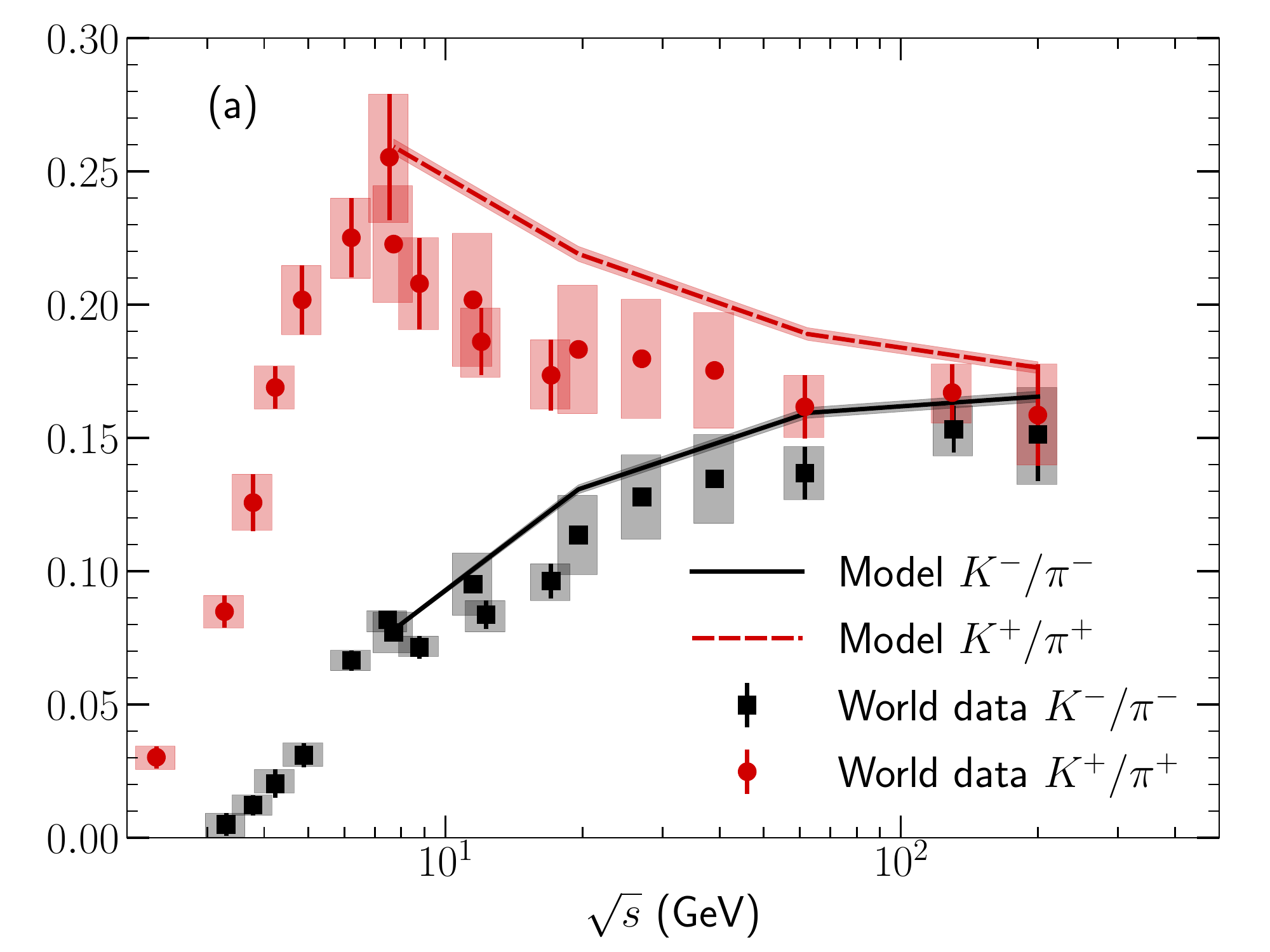}  & \includegraphics[width=0.4\linewidth]{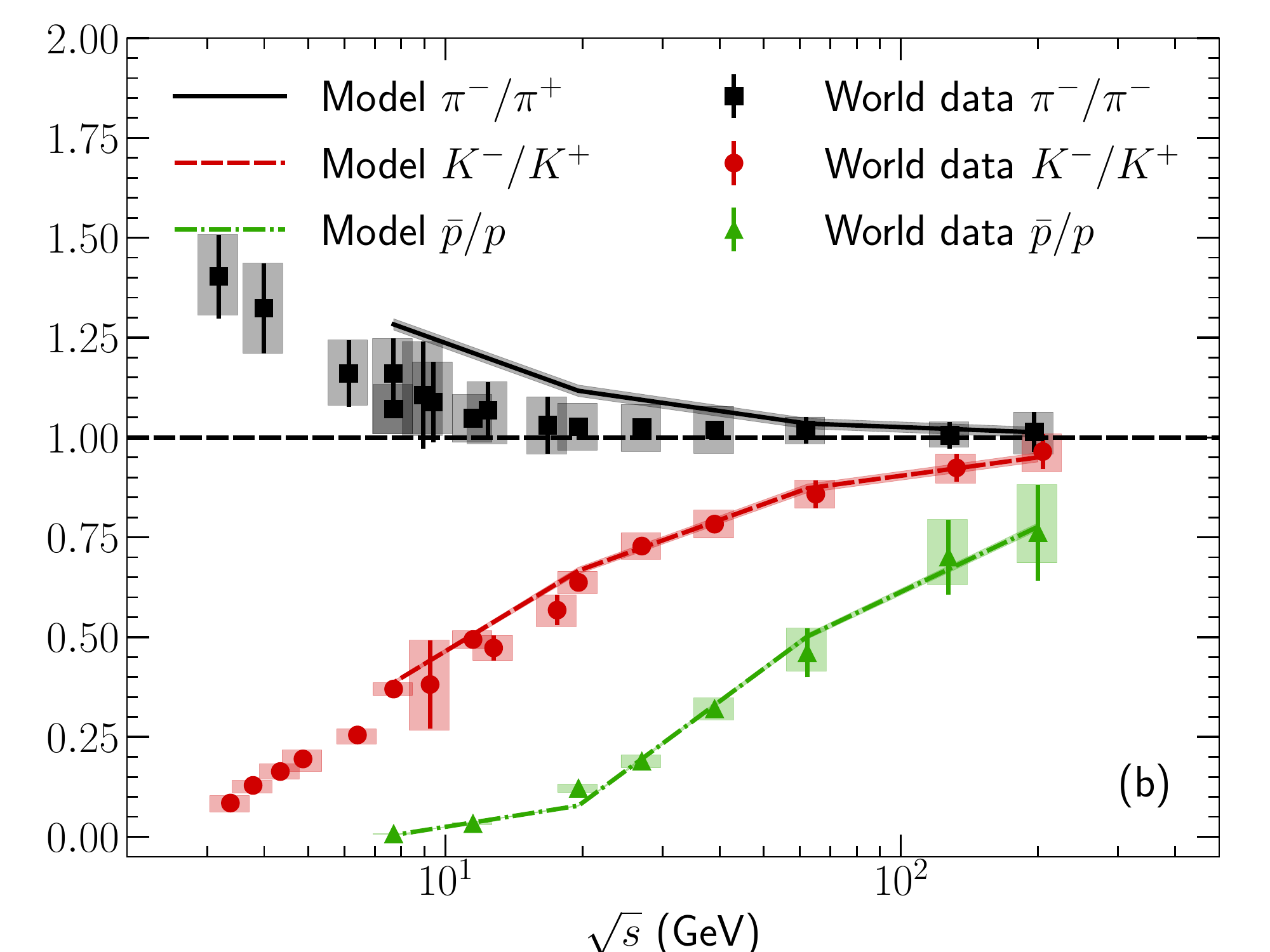} \\
   \includegraphics[width=0.4\linewidth]{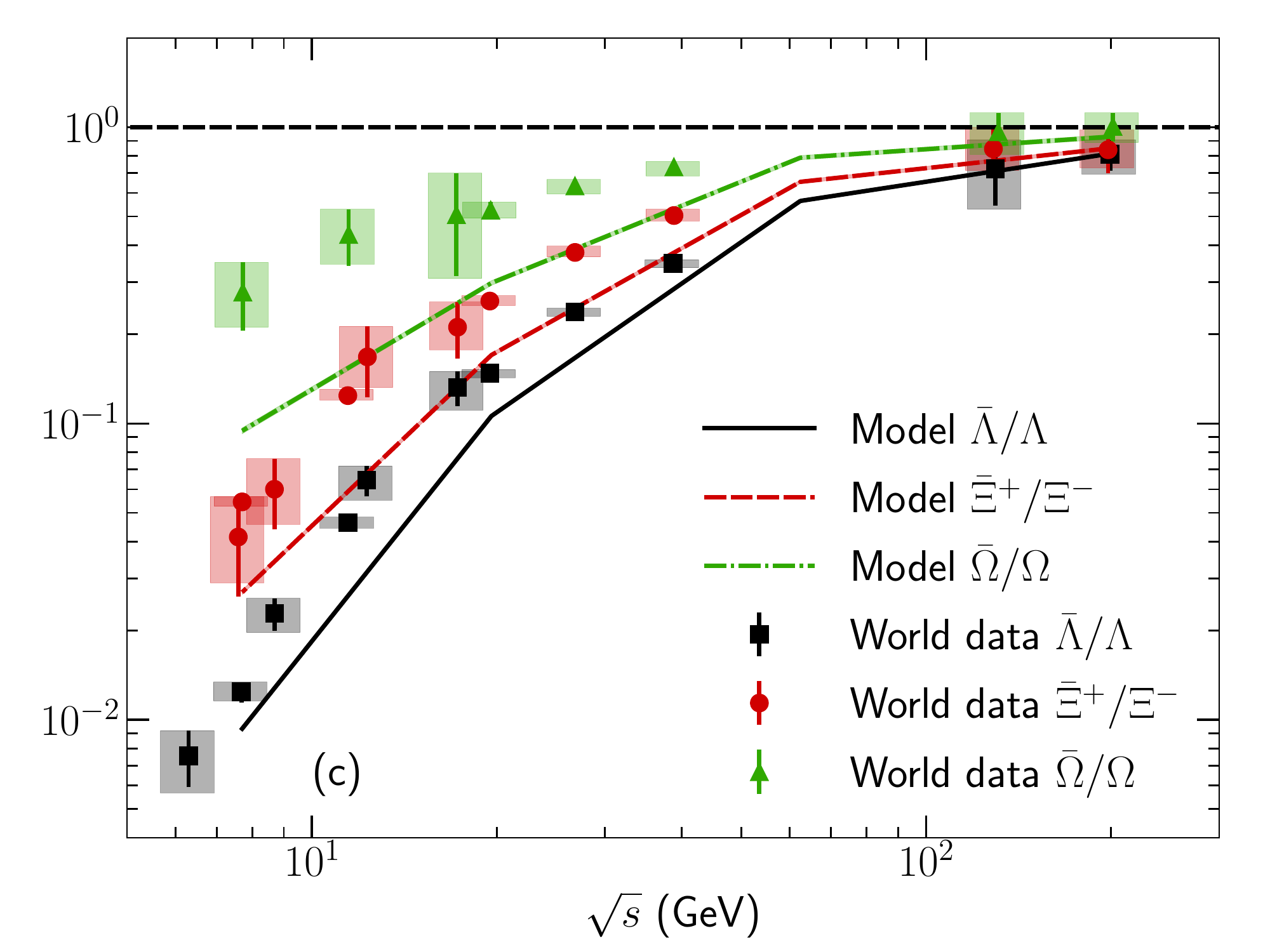} & 
   \includegraphics[width=0.4\linewidth]{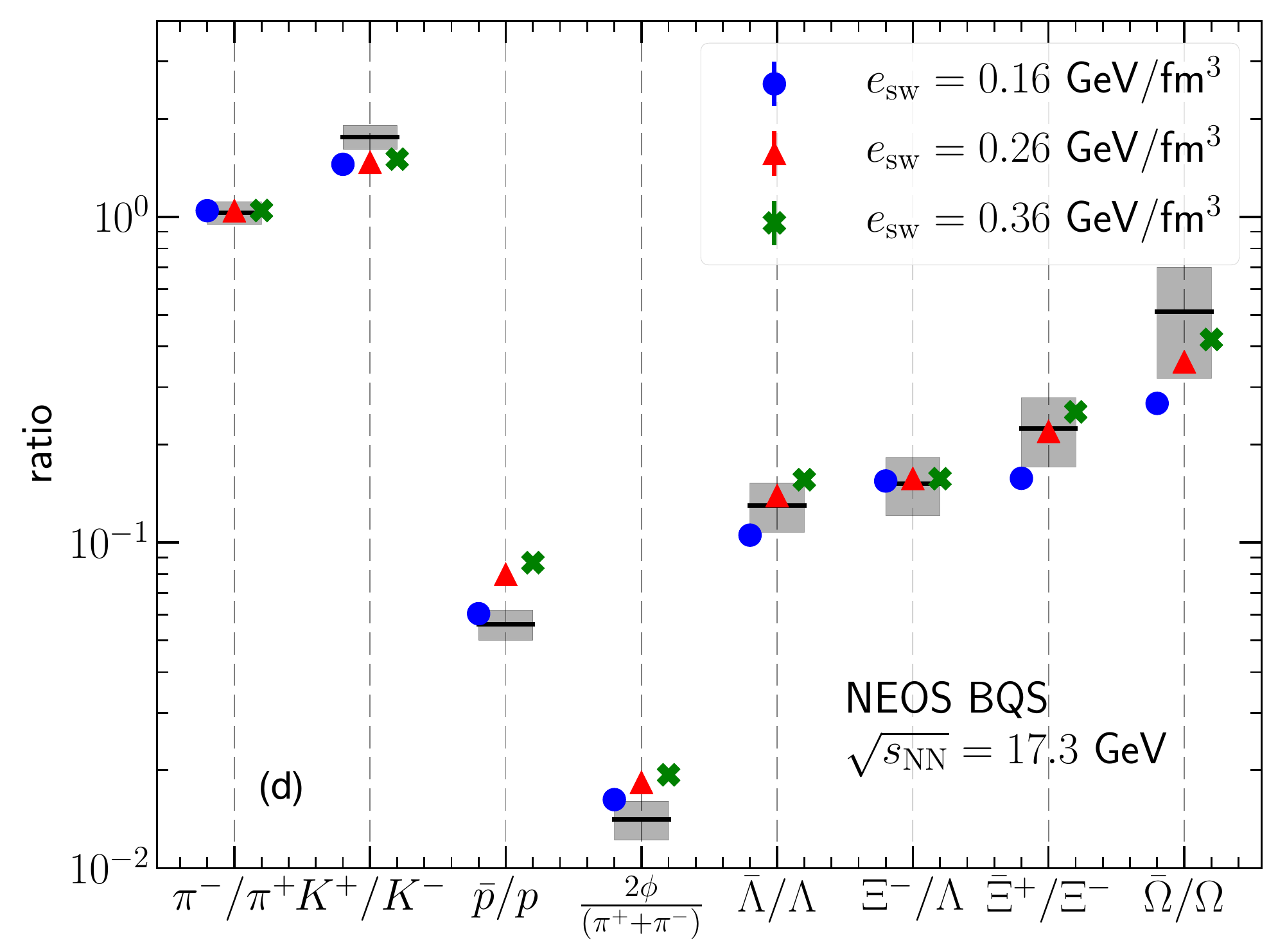}
\end{tabular}
\caption{Panels (a, b, c): Preliminary calculations of identified particle ratios as functions of the collision energy in central Au+Au collisions compared with the world data \cite{E895:2001zms, NA49:2002pzu, NA49:2004irs, NA49:2008ysv, STAR:2017sal, STAR:2019bjj}. Panel (d): Identified particle ratios at three switching energy densities for central Pb+Pb collisions at the top SPS energy \cite{Monnai:2019hkn}.}
\label{fig:BES_pid_ratios}
\end{figure}
Figure~\ref{fig:BES_pid_ratios} shows the comparison between the preliminary calculations from the 3D string deceleration model \cite{Shen:2017bsr} and the identified particle yield ratios in central Au+Au collisions at the RHIC BES energies. The model calculations with a constant switching energy density $e_\mathrm{sw} = 0.26$ GeV/fm$^3$ reproduce the collision energy dependence of the data from 7.7 GeV to 200 GeV. The key theoretical ingredients that lead to this good description are initial-state baryon stopping plus the constraints $n_S = 0$ and $n_Q = 0.4 n_B$ for Au+Au collision \cite{Monnai:2019hkn, Monnai:2021kgu}. The strangeness neutrality condition $n_S = 0$ results in a linear correlation between the strangeness chemical potential and $\mu_B$, $\mu_S \sim \mu_B/3$ \cite{Monnai:2021kgu}. The collision energy dependence of $\mu_S$ can be constrained with the $K^+/\pi^+$ and $K^-/\pi^-$ ratios. Below $\sqrt{s_\mathrm{NN}} = 5$ GeV, the $K^+/\pi^+$ ratio becomes smaller because of the canonical suppression of the strange quark production \cite{Andronic:2017pug, STAR:2021hyx}.
The small $\mu_Q$ from the constraint $n_Q = 0.4 n_B$ quantitatively reproduces the $\pi^-/\pi^+$ ratios.
The model calculation underestimates the ratios among multi-strangeness baryons for $\sqrt{s_\mathrm{NN}} < 40$ GeV, which hints that these strangeness baryons could have a higher chemical freeze-out energy density than the non-strange hadrons \cite{Bellwied:2018tkc} illustrated in Fig.~\ref{fig:BES_pid_ratios}d.

\section{Outlook and challenges}

Over the past decade, the RHIC Beam Energy Scan program has excited a wave of theory developments towards a (3+1)D paradigm of dynamical modeling of relativistic nuclear collisions. The advancements in 3D initial-state models, equation of state at finite net baryon density, dynamical initialization schemes enable us to quantify initial baryon stopping and study the collectivity of the QGP in a baryon-rich environment. To confront precision flow measurements from the upcoming RHIC BES program phase II, our community has adopted the Bayesian Inference method as a standard approach to systematically constrain the QGP's thermodynamic and transport properties and initial-state fluctuation spectrum \cite{Bernhard:2019bmu, JETSCAPE:2020mzn, Nijs:2020roc}. Meanwhile, the increasing theoretical and numerical complexities in 3D hybrid models could pose significant challenges in future high-dimensional Bayesian analysis.

The current RHIC fixed-target experiment and the future FAIR/NICA experimental programs push the exploration of the QCD phase diagram beyond $\mu_B = 400$ MeV, where guidance from lattice QCD on EoS is very limited. Searching for experimental signals of critical point and first-order phase transition requires the theoretical frameworks to be calibrated by particle production and flow measurements to provide reliable baseline expectations without critical phenomena \cite{Zhao:2020irc, Oliinychenko:2020znl, Pratt:2020ekp, Vovchenko:2021kxx}. Further studying the effects of out-of-equilibrium evolution of critical fluctuations \cite{Rajagopal:2019xwg, Du:2020bxp, Dore:2020jye, Nahrgang:2020yxm, Du:2021zqz} on net proton high-order cumulants and light nuclei productions \cite{Sun:2017xrx} as functions of collision energy and rapidity intervals would shed light on the QCD phase structure at large baryon density.

\section*{Acknowledgments}
This work is support in part by the National Science Foundation (NSF) under grant number PHY-2012922 and by the U.S. Department of Energy, Office of Science, Office of Nuclear Physics, under contract number DE-SC001346 and within the framework of the Beam Energy Scan Theory (BEST) Topical Collaboration. The calculations in this work used resources provided by the Open Science Grid~\cite{Pordes:2007zzb, Sfiligoi:2009cct}, which is supported by the NSF award 2030508.

\bibliography{dynamicalModels}

\end{document}